\documentclass[12pt]{JHEP3}
\usepackage{amsfonts,amssymb}
\usepackage{amsmath}

\usepackage{epsf}

\newcommand{\C}[1]{\mathbb{C}^{#1}}
\newcommand{\R}[1]{\mathbb{R}^{#1}}

\def\Z{\mathbb{Z}}
\def\N{\mathbf{N}}

\def\T11{{T}^{1,1}}
\def\bear{\begin{eqnarray}}
\def\eear{\end{eqnarray}}

\def\bX{\mathbf{X}}
\def\bY{\mathbf{Y}}

\def\vol{\mathrm{vol}}

\newcommand{\pa}{\partial}
\newcommand{\tr}{{\rm tr}}
\newcommand{\comment}[1]{}

\newcommand{\pasl}{\pa\kern-.55em /}

\newcommand{\ksl}{k\kern-.55em /}

\DeclareFixedFont{\xiiss}{OT1}{cmss}{m}{n}{12}
\DeclareFixedFont{\ixss}{OT1}{cmss}{m}{n}{9}
\DeclareFixedFont{\cmrnine}{OT1}{cmr}{m}{n}{9}
\newcommand{\field}[1]{\mathbb{#1}}

\newcommand{\BC}{{\field C}}

\newcommand{\BZ}{{\field Z}}

\newcommand{\CCs}{\hbox{\ixss C\kern-.4emI}}
\newcommand{\ZZs}{\hbox{\ixss Z\kern-.4emZ}}

\newcommand{\CP}{{\BC\field P}}

\title{Baryon Spectra and AdS/CFT Correspondence}
\author{David Berenstein\\ 
School of Natural Sciences, Institute for Advanced Study, Einstein Drive,
Princeton, NJ  08540, USA\\
Email: \email{dberens@ias.edu}}
\author{Christopher P. Herzog and Igor R. Klebanov\\
Joseph Henry Laboratories, Princeton University, Princeton, NJ  08544, USA\\
Email: \email{cpherzog@princeton.edu}, \email{klebanov@feynman.princeton.edu}}
\abstract{ 
We provide a detailed map between wrapped D3-branes
in Anti-de Sitter (AdS) backgrounds 
and dibaryon operators in 
the corresponding conformal field theory (CFT). 
The effective five dimensional action governing
the dynamics of AdS space contains a $U(1)_R$ gauge field
that mediates interactions between objects possessing R-charge. 
We show that the $U(1)_R$ charge of these wrapped 
D3-branes as measured by the  gauge field matches the R-charge
of the dibaryons expected from field theory considerations.  
We are able, through a careful probe brane calculation
in an $AdS_5\times T^{1,1}$ background, 
to understand the exact relation between the mass of the 
wrapped D3-brane and the dimension of the corresponding dibaryon.
We also make some steps
toward matching the counting of dibaryon operators in the CFT with the
ground states of a supersymmetric quantum mechanical system
 whose target space is the moduli space of D-branes. 
Finally, we discuss BPS excitations of the D3-brane and 
compare them with higher dimension operators in the CFT.   
}

\preprint{hep-th/0202150\\
PUPT-2023}

\begin{document}

\section{Introduction}

According to the AdS/CFT conjecture \cite{jthroat,GKM,EW}, the chiral operators
of the ${\cal N}=4$ supersymmetric $SU(N)$ gauge theory are in
one-to-one correspondence with the modes of type IIB supergravity
on $AdS_5 \times S^5$. 
On the other hand, the massive string modes correspond to
operators in long multiplets whose dimensions diverge 
for large `t Hooft coupling as $(g_{\rm YM}^2 N)^{1/4}$.
Thus, in the limit $g_{\rm YM}^2 N\rightarrow \infty$
the stringy nature of the dual theory is obscured by the
decoupling of the non-chiral operators, which constitute the large
majority of possible gauge invariant operators. 

It is important to keep in mind, however, that the AdS/CFT correspondence
relates large $N$ gauge theory to string theory, not merely supergravity.
If we depart from the limit of infinite `t Hooft coupling,
then all the non-chiral operators no longer decouple, so that
the spectrum of the gauge theory becomes much more complicated and presumably 
related to type IIB string theory on $AdS_5\times S^5$ (see, for example,
\cite{Polyakov}). However, even for very large
`t Hooft coupling, it is possible to demonstrate the stringy nature
of the dual theory quite explicitly. One early clue was provided by
Witten \cite{WSO} in the context of 
${\cal N}=4$ supersymmetric $SO(2N)$ gauge theory which is dual to
type IIB strings on $AdS_5\times RP_5$. He noted that the gauge theory
possesses chiral operators of dimension $N$, the Pfaffians, whose
dual is provided by a D3-brane wrapping a 3-cycle of the $RP_5$.
This explicitly shows that the dual theory cannot be simply
supergravity: it must contain D-branes. Since 
the third homology class $H_3(RP_5)= \Z_2$,
a D3-brane wrapped twice can decay into ordinary supergravity modes.
A corresponding gauge theory statement is that a 
product of two Pfaffian operators can be expressed in terms of the usual
single-trace operators.

It is further possible to find type IIB backgrounds of the form
$AdS_5\times \bX_5$ such that $H_3(\bX_5)= \Z$. In this case, a D3-brane
can wrap a 3-cycle any number of times so that the dual gauge theory operators
carry a quantized baryon number. A simple example of this kind is provided
by the space $\bX_5 = T^{1,1}= (SU(2)\times SU(2))/U(1)$
whose symmetries are $U(1)_R\times SU(2)\times SU(2)$.
This space may be thought of as a $U(1)$ bundle over $S^2\times S^2$, and the 
explicit Einstein metric on $T^{1,1}$ is
\begin{equation} \label{co}
d s_5^2=
{\frac 1 9} \bigg(d\psi + 
\sum_{i=1}^2 \cos \theta_i d\phi_i\bigg)^2+
{\frac 1 6} \sum_{i=1}^2 \left(
d\theta_i^2 + {\rm sin}^2\theta_i d\phi_i^2
 \right)
\ .
\end{equation}

The ${\cal N}=1$
superconformal gauge theory dual to type IIB strings on
$AdS_5\times T^{1,1}$ was constructed in \cite{KW}:
it is $SU(N)\times SU(N)$ gauge theory
coupled to two chiral superfields, $A_i$, in the $({\bf N}, \overline{\bf N})$
representation
and two chiral superfields, $B_j$, in the $(\overline{\bf N}, {\bf N})$
representation \cite{KW}. The $A$'s transform as a doublet under one
of the global $SU(2)$'s while the $B$'s transform
as a doublet under the other $SU(2)$.

Cancellation of the anomaly in the $U(1)$ R-symmetry requires that
the $A$'s and the $B$'s each have R-charge $1/2$. For consistency of
the duality it is necessary that we add
an exactly marginal superpotential which preserves the
$SU(2)\times SU(2)\times U(1)_R$ symmetry of the theory:
\begin{equation}
W=\epsilon^{ij}
\epsilon^{kl}\tr A_iB_kA_jB_l\ .
\end{equation}

In \cite{GK} it was proposed that the  
wrapped
D3-branes correspond to baryon-like operators $A^N$
and $B^N$ where the indices of both $SU(N)$ groups are fully
antisymmetrized (their more detailed form is exhibited in Section 3).
For large $N$ the dimensions of such operators
calculated from the mass of wrapped D3-branes 
were found to be $3N/4$ \cite{GK}. 
This is in
complete agreement with the fact that the dimension of the chiral
superfields at the fixed point is $3/4$.
At the quantum level, the collective
coordinate for the wrapped D3-brane has to be quantized to identify
its $SU(2)\times SU(2)$ quantum numbers. 
This quantization was sketched in \cite{GK}. Here we do a more careful
job and show that the dimension of dibaryons is $3N/4$ not only in
the large $N$ limit but exactly. 
In a related check on the identification of the 
dibaryons with wrapped D3-branes,
we calculate the $U(1)_R$ charge of the wrapped D3-branes and show
that it is $N/2$, in agreement with the gauge theory.       

The structure of the paper is as follows. In 
Section 2 we calculate the $U(1)_R$ charge of a D3-brane wrapping
a 3-cycle inside the Einstein space $\bX_5$. Our results apply to all
$\bX_5$ which are $U(1)$ bundles over 4-d K\"ahler-Einstein spaces.
We also provide an analogous calculation for M5-branes wrapping
a 5-cycle inside an Einstein space $\bX_7$ which is a $U(1)$ bundle
over a 6-d K\"ahler-Einstein space. In all cases, we show that
the R-charge is proportional to the dimension $\Delta$ of the
dual CFT operator, with the correct constant of proportionality.
We further argue that $\Delta$ is measured by the volume of the cycle wrapped
by the brane. 
In Section 3 we make some remarks on the collective coordinate
quantization of the wrapped D3-branes.  In particular,
we focus on the $AdS_5 \times T^{1,1}$ example and show that the $SU(2)\times
SU(2)$ quantum numbers of the dibaryon operators indeed follow from this
quantization.
In Section 4 
we study BPS excitations of the wrapped D3-branes
using the DBI action. We calculate their energies and $U(1)_R$ charges
and propose dual chiral operators carrying the same quantum numbers.

\section{The R-symmetry gauge field}

We would like to show how the $U(1)_R$ gauge field emerges 
from a Kaluza-Klein reduction of the full ten or eleven 
dimensional supergravity 
action.\footnote{
This gauge field is also discussed in \cite{KPW}
where the presence of fractional branes causes
the field to acquire a mass.}
In ten dimensions, we begin with a stack of D3-branes placed at the tip
of a non-compact Calabi-Yau cone $\bY_3$ of complex dimension three.
In eleven dimensions, we consider a stack of M2-branes at the tip of
a similar cone $\bY_4$ of complex dimension 4.  
We may write the metric on $\bY_n$ as 

\begin{equation}
ds_{\bY_n}^2 = dr^2 + r^2 ds_{\bX_{2n-1}}^2 \ ,
\label{Ymetric}
\end{equation}
where $\bX_{2n-1}$ is Einstein, i.e.~$R_{ab} = 2(n-1) g_{ab}$.

In the near-throat limit \cite{jthroat}, the backreaction of the
branes causes the full space to separate into a product manifold.
In the D3-brane case, we have $AdS_5\times \bX_5$ and
a five form flux $dC_4 = F_5$ produced by the stack of D3-branes. 
See also \cite{MP} for details.  
For the M2-branes, the full space becomes $AdS_4\times \bX_7$,
and the M2-brane flux is carried by $dC_3 = F_4$.
 
When $\bX_5$ is $S^5$ or the coset manifold
$T^{1,1}$, the authors of 
\cite{KRN, Ceresole} demonstrate that perturbations
in the metric $h_{\mu i}$ combine with perturbations in the
$C_{\mu i j k}$ potential to produce a massless vector field.
(The $i,j, \ldots$ index $\bX_m$ while $\mu,\nu,\ldots$ index
$AdS$ space.)  For $\bX_7$, more general results exist concerning
the existence of this massless vector field \cite{DAuria}.
In particular, whenever $\bX_7$ is a coset space and there
exists a Killing vector, harmonic analysis demonstrates the
existence of a massless vector field mixing perturbations 
of $h_{\mu i}$ with perturbations to $C_{\mu i j}$.

In what follows, we consider a more general class of $\bX_n$,
in particular quasi-regular Einstein-Sasaki manifolds,
and demonstrate the existence of a massless U(1) vector
field.  We argue that this massless vector field
mediates the interaction between objects with R-charge.
Although the generalization is interesting in its own
right, the main interest is the demonstration that
wrapped D3-branes and M2-branes in these geometries
always carry an R-charge consistent with their
identification 
as dibaryons in the corresponding conformal 
field theory.

To be more specific, 
a quasi-regular Einstein-Sasaki manifold is essentially
a U(1) fibration of a K\"ahler-Einstein manifold or orbifold.
 The BPS dibaryons wrap the base of the Sasaki-Einstein manifold
in a holomorphic manner, and they wrap the $S^1$ fiber completely.
Mikhailov has considered similar holomorphically wrapped 
D-branes \cite{Mik}.
One might naively say that
the R-charge of a particle is its angular momentum along the 
$S^1$ direction. Thus, a brane that wraps the fiber 
carries no net angular momentum. 
However, one must be careful; these branes can be 
supersymmetric,
and it is necessary to understand what the precise notion of R-charge 
is in the supergravity.
The following calculation
is not that sensitive to the dimensionality, but for clarity,
we will consider the $AdS_4$ and $AdS_5$ cases separately.

\subsection{The R-charge for $AdS_5 \times \bX_5$}

In general, the ten dimensional metric is
\begin{equation}
ds^2 = \frac{r^2}{L^2}\eta_{\alpha \beta}dx^\alpha dx^\beta + 
L^2 \frac{dr^2}{r^2} +
L^2 ds^2_{\bX_{5}} , 
\end{equation}
where $\eta_{\alpha \beta}=(-+++)$ is a Minkowski tensor.  

A solution to the SUGRA equations of motion can be obtained by 
threading the $\bX_{5}$ space with $N$ units of $F_{5}$ form
flux.  In particular
\begin{equation}
F_5 = {\mathcal F} + \star{\mathcal F} \; ; \; \; \;
{\mathcal F} = 4 L^{4} \vol(\bX_{5})
\end{equation}
where $L^4 \sim g_s N$.  

We now further specialize to the case where $\bX_5$ is a 
quasi-regular Einstein-Sasaki space.  In simpler language, 
$\bX_5$ is a U(1) bundle over a two complex dimensional 
K\"ahler-Einstein manifold (or orbifold) $V$:
\begin{equation}
ds_{\bX_5}^2 = \left(\frac{q}{3} \right)^2 
\left(d\psi + \frac{3}{q} \, \sigma \right)^2 + 
h_{\alpha \bar\beta} dz^\alpha d\bar{z}^\beta \ .
\end{equation}
Define the K\"ahler form on $V$ to be
$\omega = ih_{\alpha \bar\beta} dz^\alpha \wedge d\bar{z}^\beta$.
Then, $d\sigma = 2\omega$.  The number $q$ is defined
such that $n \, d\sigma = 2\pi q c_1$, where $c_1$ is the
first Chern class of the U(1) bundle and $n=3$.  With these definitions,
$0 \leq \psi < 2\pi$ \cite{BH}.
Also, in this way, we
may write
\begin{equation}
\vol(\bX_5) = \frac{q}{3!} \, \eta \wedge \omega^{2}
\end{equation} 
where $\eta = d\psi + \frac{3}{q} \sigma$, and 
$\omega \wedge \omega = \omega^2$.

We want to identify the gauge field $A$ associated with the R-charge.
The R-symmetry gauge group is at least as big as U(1), and hence it
is a natural guess to associate the 
U(1) fiber of the Sasaki-Einstein manifold
$\bX_5$ if not with the R-symmetry group itself, then at least with
a subgroup of it.  We begin with the inspired guess that redefinitions of
$\psi$ be identified with gauge transformations of the R-symmetry gauge field
$A$:
\begin{equation}
\psi \rightarrow \psi + \alpha \epsilon
\end{equation}
where
\begin{equation}
A \rightarrow A + d\epsilon \ .
\end{equation}
One troublesome issue is that $\alpha$ in general may not be one.
To figure out $\alpha$ and to anchor the analysis, 
we need a geometric object with a 
definite R-charge.  The holomorphic $n$-form $\Omega$ on the noncompact
Calabi-Yau cone $\bY_n$
is one such object.  Because of its association with the superpotential, 
$\Omega$ has R-charge 2.  

The functional dependence of $\Omega$ on $\psi$ is very simple: 
$\Omega \sim \exp(i q \psi)$.  A proof will follow in section 2.3.  
Because
$\Omega$ has R-charge two, we can directly 
identify the angular variable $\psi_R$ with
a U(1) R-symmetry where now
\begin{equation}
\psi_R \rightarrow \psi_R + \epsilon
\end{equation}
and $2\psi_R = q \psi$.  Thus $\alpha = 2/q$.

Including the gauge field $F$ in the supergravity solution means altering
the metric 
\begin{equation}
ds_{\bX_5}^2 = \left(\frac{q}{3} \right)^2 
\left(d\psi + \frac{3}{q} \, \sigma + \frac{2}{q} A \right)^2 + 
h_{\alpha \bar\beta} dz^\alpha d\bar{z}^\beta \ .
\label{modmetric}
\end{equation}
The $F_5$ form flux is also altered
\begin{equation}
{\mathcal F} = 4 L^{4} 
\frac{q}{3!} \left((\eta+\frac{2}{q}A)\wedge \omega^2-
\frac{1}{3} \, dA \wedge \eta \wedge \omega \right) \ .
\end{equation}
{}From this expression, we can calculate $C_4$ to first order in $A$:
\begin{equation}
C_{4} = 4 L^{4} \frac{q}{3!} 
\left( \psi \, \omega^2 - 
\frac{1}{3} \, A \wedge \eta \wedge \omega \right) 
- \frac{r^4}{L^4} d^4x + \frac{q}{9}L^3(\star_5 dA) \wedge \eta \ .
\end{equation}
We have introduced some notation: 
$d^4x = dx^0\wedge dx^1 \wedge dx^2 \wedge dx^3$, and $\star_5$
is the Hodge star in the $AdS_5$ directions.
This form of $F_{5}$ satisfies the SUGRA equations of motion
to linear order in $A$.  In particular, $F_5$ has been 
constructed such that $d{\mathcal F}=0$.  
Notice that $\star {\mathcal F}$ contains a piece
proportional to $(\star_5 dA) \wedge \omega$.  Hence
$d\star {\mathcal F}$ will vanish to linear order 
in $A$ provided $d \star_5 dA = 0$, i.e. the field
equations for the gauge field $F=dA$ are satisfied.

We also may consider the trace of the Einstein
equations.  The modification of the metric (\ref{modmetric}) 
causes the Ricci scalar for the full ten dimensional
metric to become
\begin{equation}
R \rightarrow R + \frac{L^2}{9} |F|^2 \ .
\end{equation}
Thus, the trace of Einstein's
equations will contain an extra $|F|^2$ term. 
This term causes the backreaction of the $U(1)_R$ field strength
on the metric. 
We ignore it in this paper since it is a second order effect.


As a test of our proposal for the U(1) R-symmetry gauge field,
let us calculate the R-charge of a hypothetical baryon in this
geometry.  We identify the baryons as D3-branes wrapped
on 3-cycles inside $\bX_5$.  
In particular, the D3-brane will completely wrap the U(1) fiber
of the Einstein-Sasaki manifold $\bX_5$
and will also wrap a holomorphic curve in the K\"ahler-Einstein base $V$.
The dimension 
of these baryons is proportional to the volume
of the 3-cycle $C$ times the curvature length $L$:
\begin{equation}
\Delta = 
\mu_3 L^4 \int_C \frac{q}{3} \eta \wedge \omega \ .
\label{Delta}
\end{equation}
One might expect that the volume is more closely related to the mass of
the baryon than to the dimension.\footnote{
It is often possible to evaluate (\ref{Delta}).  The integral
factors into two pieces: the length of the U(1) fiber and 
an integral over a holomorphic curve.  The length of
the U(1) fiber is trivial, $2\pi$.  Using the
Einstein condition, $\omega$ can be related to the
first Chern class of $V$, $c_1(V)$.  The second 
integral becomes an intersection number calculation.
}
We will see later in section 2.4
that the mass receives a correction and that $(\ref{Delta})$ is
the correct expression.

Now these baryons are thought to be chiral primary operators.  Their
R-charge should be a multiplicative constant times the dimension.
Recall that in a classical Lagrangian for a particle traveling in
an electromagnetic field, there is a term of the form
$q A_\mu v^\mu$ where $q$ is the charge of the particle and $v^\mu$
is its four-velocity.  
In the Lagrangian for a probe baryon of this type, there is 
a Wess-Zumino term of the form
\begin{equation}
\mu_3 \int_{C} C_4 = \Delta \frac{2}{3} A \ ,
\end{equation}
where we have used (\ref{Delta}) for $\Delta$.

In the case of $AdS_5 \times \bX_5$, we know independently
from the supersymmetry algebra that 
the dimension times 2/3 is the R-charge.
The $2\Delta/3$ in the Wess-Zumino term is exactly where the 
charge should be in a classical Lagrangian describing
the D3-brane dynamics.

\subsection{The R-charge for $AdS_4 \times \bX_7$}

We now repeat essentially the same analysis for 
a stack of M2-branes sitting at the 
tip of a noncompact four complex dimensional cone,
$\bY_4$.
In general, the eleven dimensional metric is
\begin{equation}
ds^2 = \frac{r^2}{L^2}\eta_{\alpha \beta}dx^\alpha dx^\beta + 
L^2 \frac{dr^2}{r^2} +
4L^2 ds^2_{\bX_{7}} , 
\end{equation}
where $\eta_{\alpha \beta}=(-++)$ is a Minkowski tensor.  Note
the additional factors of two scaling the Einstein metric.  These
factors are necessary to guarantee that the Ricci scalar of the
eleven dimensional metric vanishes and will be crucial later
on in fixing the R-charge of the baryons.

A solution to the SUGRA equations of motion can be obtained by 
threading the $\bX_{7}$ space with $N$ units of $F_{7}$ form
flux.  In particular
\begin{equation}
F_7 = 6 (2L)^{6} \vol(\bX_7)
\end{equation}
where $L^6 \sim N$.  

We specialize to the case where $\bX_7$ is a 
quasi-regular Einstein-Sasaki space:
\begin{equation}
ds_{\bX_7}^2 = \left(\frac{q}{4} \right)^2 
\left(d\psi + \frac{4}{q} \, \sigma \right)^2 + 
h_{\alpha \bar\beta} dz^\alpha d\bar{z}^\beta \ .
\end{equation}
We keep the same definition of $q$ and $\sigma$.  
Namely, $d\sigma = 2\omega$ and the number $q$ is defined
such that $n \, d\sigma = 2\pi q c_1$ and $n=4$.  With these definitions,
$0 \leq \psi < 2\pi$.
Also, in this way, we
may write
\begin{equation}
\vol(\bX_7) = \frac{q}{4!} \, \eta \wedge \omega^{3}
\end{equation} 
where $\eta = d\psi + \frac{4}{q} \sigma$.

Just as in the previous section, redefinitions of
$\psi$ can be identified with gauge transformations of the R-symmetry gauge field $A$:
\begin{equation}
\psi \rightarrow \psi + \frac{2}{q} \epsilon
\end{equation}
where
\begin{equation}
A \rightarrow A + d\epsilon \ .
\end{equation}

Including the gauge field $F$ in the supergravity solution changes
the metric 
\begin{equation}
ds_{\bX_7}^2 = \left(\frac{q}{4} \right)^2 
\left(d\psi + \frac{4}{q} \, \sigma + \frac{2}{q} A \right)^2 + 
h_{\alpha \bar\beta} dz^\alpha d\bar{z}^\beta \ .
\label{modmetric2}
\end{equation}
The $C_6$ form potential becomes
\begin{equation}
C_{6} = 6 (2L)^{6} \frac{q}{4!} 
\left( \psi \, \omega^3 - 
\frac{1}{4} \, A \wedge \eta \wedge \omega^2 \right) \ .
\end{equation}
Hence 
\begin{equation}
F_7 = 6 (2L)^{6} 
\frac{q}{4!} \left((\eta+\frac{2}{q}A)\wedge \omega^3-
\frac{1}{4} \, dA \wedge \eta \wedge \omega^2 \right) \ .
\end{equation}
This form of $F_{7}$ satisfies the SUGRA equations of motion
to linear order in $A$.  The relevant equations for $F_7$ are
$dF_7 = \frac{1}{2}(\star F_7) \wedge (\star F_7)$ and $d\star F_7 =0$.
We have
constructed $F_7$ such that $dF_7=0$.  Moreover,
$(\star F_7) \wedge (\star F_7)$ and $d\star F_7$ 
vanish to order
$A^2$ provided the equations of motion
for the gauge field $F=dA$ are satisfied, i.e.
$d\star_4 dA = 0$.
The modification of the metric (\ref{modmetric2}) 
causes the Ricci scalar to become
\begin{equation}
R \rightarrow R + \frac{L^2}{4} |F|^2 \ .
\end{equation}
{}From the standpoint of an effective action in $AdS_4$,
one can see from this modification another way of
deriving the equations of motion for the gauge field,
$d\star_4 dA = 0$.


To calculate the R-charge of a hypothetical baryon in this
geometry, we identify the baryons as M5-branes wrapped
on 5-cycles inside $\bX_7$.  The 5-cycle consists
of the U(1) fiber and a holomorphic surface inside
the K\"ahler-Einstein base of $\bX_7$.
The dimension 
of these baryons is proportional to the volume
of the 5-cycle $C$ times the curvature length $L$:
\begin{equation}
\Delta = 
\mu_5 (2L)^5L \int_C \frac{q}{8} \eta \wedge \omega^2 \ .
\label{Delta2}
\end{equation}
In the Lagrangian for a probe baryon of this type, there is also 
a Wess-Zumino term of the form
\begin{equation}
\mu_5 \int_{C} C_6 = \Delta A \ ,
\end{equation}
where we have used the previous expression for $\Delta$.

In the case of $AdS_4 \times \bX_7$, we expect from 
the supersymmetry algebra that the R-charge 
of a chiral primary operator be equal
to the dimension.  
As these baryons are chiral primary operators, all is well.

\subsection{Proof}

We seek to prove that the holomorphic $n$-form $\Omega$
scales as $\exp{iq\psi}$ where $\psi$ is a coordinate on
the U(1) fiber of these non-compact Calabi-Yau cones of
complex dimension $n$.
The metric on $\bY_n$ (\ref{Ymetric}) is not very natural 
as it puts $r$ and $\psi$
on an unequal footing.  It is more natural to think of $\bY_n$ as a 
fibration of $\C{*}$ over the K\"ahler-Einstein base $V$.  Let
$\lambda \in \C{*}$ be the coordinate of the fiber.  We
make the change of variables
\begin{equation}
\lambda = r^{n/q} e^{i\psi} \ .
\end{equation}
In these coordinates, the metric becomes
\begin{equation}
ds_{\bY_n}^2 = \left(\frac{q}{n}\right)^2 
(\lambda \bar\lambda)^{q/n - 1} (d\lambda+i\lambda\sigma)
(d\bar\lambda -i \bar\lambda \sigma)
+ (\lambda \bar\lambda)^{q/n} h_{\alpha \bar\beta} dz^\alpha d\bar{z}^\beta
\end{equation}

The volume form scales as $r^{2n} = (\lambda \bar\lambda)^{q}$.
Note that the volume form is proportional to $\Omega \wedge \bar\Omega$.
Thus, $\Omega$ scales as $\lambda^{q}$.

\subsection{The difference between mass and dimension}

So far we have explicitly shown that the supergravity vector field 
responsible for the $U(1)_R$ charge of a state results from mixing 
between the rotations along the $S^1$ fiber and a compensating gauge 
transformation. The end result is that the D-brane 
R-charge was identified with the volume of the D-brane 
for certain BPS D-branes.

In the supergravity, it seems natural to associate to such a D-brane 
wrapping the given cycle a mass equal to the volume, and then one
finds a discrepancy between the R-charge and the conformal dimension, 
since
\begin{equation}\label{eq:confdim}
\Delta = d/2 +\sqrt{m^2 L^2 + d^2/4}
\end{equation}
for $AdS_{d+1}$.
We argue that the volume of the D-brane should be identified
directly with the conformal dimension, and not with the mass of the
particle.

One reason for this is that the D-brane is a BPS object and we should
consider the action of a superparticle in this background, rather
than the action of a point particle in AdS space.
When we consider global AdS coordinates, the hamiltonian associated to
 time corresponds to the generator of conformal tranformations, and
 the spectrum is discrete because there is a gravitational
potential well at the center of AdS which eliminates the zero modes
of the particle and turns them into oscillators.

The supersymmetries do not commute with the hamiltonian, but they
still control the dynamics of the theory. This can be seen explicitly
in the matrix model presented in \cite{BMN},
 where the fermions and bosons are not degenerate, but the
ground state energy still cancels. We argue that this is generic for
these systems.
A superparticle also has fermionic coordinates. It is the zero point
 energy of these degrees of freedom that cancels the zero point energy of
the bosonic degrees of freedom.

In particular, consider the fluctuations
corresponding to the transverse motion of the particle in AdS space.
The fluctuations
have a mass term determined by the $AdS_{d+1}$ metric:
\begin{equation}
ds^2 =-  dt^2(\cosh^2 \rho) +d\rho^2 + \sinh^2 \rho d\Omega_{d-1}^2
\ .
\end{equation}
If we consider a particle at $\rho=0$, we have a timelike
geodesic of $AdS_{d+1}$.
To quadratic order in the transverse fluctuations we have
\begin{equation}
ds^2 =-  dt^2(1+\vec\rho{\,}^2) +d\vec\rho{\,}^2 \ .
\end{equation}
There is a gravitational potential in $g_{00}$, and the  mass term
for the fluctuations in $\vec \rho$ is
determined by the form of the metric.
One can also determine this mass term from the algebra of isometries.
In the radial quantization, acting with derivatives changes the
conformal weight by one unit. This is the same as acting with one of
the raising operators  for $\vec \rho$ to first order in the quantum
system, because the symmetry is non-linearly realized.

Similarly, if we act with the supersymmetry generators, we change the
conformal weight by a factor of 1/2.
This means that the fermions will have a
mass term which is one half that of the bosons. There are $d$ complex
supersymmetries, so we have one complex fermion associated to each one
of them. It is easy to see that the contribution to the conformal
weight of each fermion
is one half that of the bosons, but there are twice as many fermions.
Thus one finds that the zero point energy cancels between bosons and
fermions, and for supersymmetric states one has
\begin{equation}
\Delta = \Delta_{class} = \hbox{Vol} \ .
\end{equation}
The classical conformal weight associated to the state is equal
to the value in the quantum theory, namely, the conformal weight of the
superparticle is the volume of the D-brane.

If we integrate out the fermions, they shift the zero point energy of the
hamiltonian by $-\frac{1}{2}N_f m_f = -d/2$, where
$N_f=2d$ is the number of real fermions, $m_f=1/2$ is the mass of each of
these fermions, and the remaining $-1/2$ is the numerical
factor associated with zero point energy for a fermionic oscillator.
As a result, the mass of the particle
associated to the ground states is given approximately by
\begin{equation}
 mL = \Delta - \frac{d}{2} \ .
\end{equation}
This expression agrees with (\ref{eq:confdim}) in the limit
of large $m$ exactly as we would expect.  To first order,
integrating out the fermions reproduces the shift between
the value of the mass and the conformal weight.

\section{Moduli space of D-branes and zero-mode quantization }

Classically one often has the freedom to change
continuously
the way in which a D-brane is wrapped without changing the energy.
In this section, we will investigate how, through quantization,
 this freedom of movement
allows one to understand the number of ground states of a given dibaryon
operator in the field theory side of the correspondence.

This abstract statement concerning the freedom some dibaryon-like states
have to shift their wrapping configurations is particularly easy
to understand in the case of $AdS_5 \times T^{1,1}$.
Recall that $T^{1,1}$ is a U(1) fibration over
$S^2 \times S^2$.
The simplest dibaryon corresponds to a D3-brane wrapping
the U(1) fiber of one of the two $S^2$'s.
It is clear that, classically,
the D3-brane can move continuously along the
other $S^2$ without changing its energy.

Let us exhibit the operators of the $SU(N)\times SU(N)$
gauge theory dual to these simply wrapped D3-branes.
Since the fields $A^{\alpha}_{k\beta}$, $k=1,2$,
carry an index $\alpha$ in the $\N$ of $SU(N)_1$ and an index $\beta$
in the $\overline{\N}$ of $SU(N)_2$,
we can construct color-singlet ``dibaryon'' operators
by antisymmetrizing completely with respect to both groups:
\begin{equation}\label{BaryonOne}
{\cal B}_{1 l}= \epsilon_{\alpha_1 \ldots \alpha_N}
\epsilon^{\beta_1\ldots \beta_N} D_{l}^{k_1\ldots k_N}
\prod_{i=1}^N A^{\alpha_i}_{k_i  \beta_i}
\ ,
\end{equation}
where $D_l^{k_1\ldots k_N}$ is the completely
symmetric $SU(2)$ Clebsch-Gordon
coefficient corresponding to forming the ${\bf N+1}$ of $SU(2)$ out of $N$ {\bf 2}'s.
Thus the $SU(2)\times SU(2)$ quantum numbers of ${\mathcal B}_{1 l}$ are
$({\bf N+1}, {\bf 1})$. Similarly, we can construct ``dibaryon'' operators
which transform as $({\bf 1}, {\bf N+1})$,
\begin{equation}\label{BaryonTwo}
{\cal B}_{2 l}= \epsilon^{\alpha_1 \ldots \alpha_N}
\epsilon_{\beta_1\ldots \beta_N} D_{l}^{k_1\ldots k_N}  
\prod_{i=1}^N B^{\beta_i}_{k_i  \alpha_i} 
\ .
\end{equation}
The existence of two types
of ``dibaryon'' operators is related on the supergravity
side to the fact that the base of the $U(1)$
bundle is $S^2\times S^2$. A D3-brane can wrap either of the
two-spheres together with the $U(1)$ fiber. 

We will use a simplified notation for the dibaryon operators 
(\ref{BaryonOne})
\begin{equation}\label{eq:dibaryon}
\epsilon^1\epsilon_2(A_{i_1}, \dots , A_{i_N}) \ .
\end{equation}
The (super)subindex on the $\epsilon$ refers to the gauge group that 
the $\epsilon$ symbol carries and whether these are (super)subindices is
indicated by the position of the group label. The gauge
indices of the $A_i$ are contracted with the gauge indices of the 
$\epsilon$ tensors in the same order as they are written.

In order to account for the $SU(2)\times SU(2)$ quantum numbers
of the dibaryons, one needs to
take into account the proper quantization of the zero modes 
of the wrapped D3-brane \cite{GK}. The resulting dynamics reduces
to motion of a particle on a sphere in the presence of a magnetic field. 
The purpose of this
section is to consider the zero mode problem in a more general setting.
After discussing the conifold case, 
we will give a more general argument for other Calabi-Yau 
singularities that will show that one can count the ground states
even without exact knowledge of the metric.

First, we describe this freedom of movement some dibaryons possess
in a more precise and useful way.
If we begin with dibaryon-like states on $AdS\times \bX$, then
at each time $t$, the D-brane
worldvolume will be wrapping  a holomorphic cycle of $\bX/U(1)$, and the
volume of the D-brane will be constant, but its shape will change.
Since the volume remains constant, there is no potential on these 
directions, and the moduli space of the D-brane can be captured by the
moduli space of holomorphic submanifolds of $\bX/U(1)$ with specific 
topological quantum numbers, which describe what holomorphic cycle is being 
wrapped.  

The D-brane is a holomorphic submanifold of codimension $1$ on $\bX/U(1)$, 
so it can be 
associated to a divisor $[D]$ on $\bX/U(1)$. 
The D-brane is the zero locus of a 
global section of the associated line bundle to the divisor $[D]$.
Any linear combination of global holomorphic sections of $[D]$ is also a 
global section, and in this way the moduli space of the D-brane is 
a projective space $\CP^m$ for some $m$ (which can be zero, and then the
 curve is rigid). $m$ is given by the number of holomorphic sections
of the line bundle associated to $[D]$ minus one.

Thus, to quantize the zero modes of the D-brane, we need the effective action
of the D-brane moduli space.  We consider our action as a nonlinear
sigma model whose target space is the moduli space of the D-brane.
This action will involve a metric on $\CP^m$
(which generically has some singularities) induced from
the metric on $\bX/U(1)$.  Additionally, one can
add a line bundle on this moduli 
space, that is, 
an effective magnetic field on the target space of the sigma model.

In the example of the conifold, 
for a single dibaryon state
 $m=1$, 
we have the round metric on $\CP^1$ (this is obvious, 
since we had an $S^2$ worth of positions for the 
D-brane with the corresponding isometry group).

Now, we want to argue that the magnetic field through this moduli space
is $N$ times the hyperplane bundle on this moduli space.
Indeed, the WZ term for the D-brane action is
\begin{equation}
N \int C_{\mu\nu\rho\sigma} dx^\mu dx^\nu dx^\rho dx^\sigma
\ .\end{equation}
Three of the coordinates of the D-brane saturate three of the indices
of $C_4$, let us say $dx^\mu dx^\nu dx^\rho$, and the fourth index is saturated
by the time direction in the form
\begin{equation}
\frac{dx^\sigma}{dt} dt
\end{equation}
So the effective action has a term proportional to the velocity
 on the moduli space
of the D-brane.  Hence, this WZW term gives rise to
an effective magnetic field on the $\CP^m$.
 Notice moreover
that this magnetic field is proportional to $N$, the effective
tension of the D-brane.
For the $S^2$ case of the conifold it is easy to show that
 the normalization
of this magnetic field  is such that the flux through the $S^2$ is $N$.
Counting the number of ground states is then done by the index theorem,
and in
this case we get $N+1$. Notice that this matches exactly the counting of
 operators in
the equation \ref{eq:dibaryon}.
For the general case we find that the index calculation counts the number of
global holomorphic functions of degree $N$ on $\CP^m$. There
are
\begin{equation}
\begin{pmatrix}N+m \\ N \end{pmatrix}
\end{equation}
such global sections.

In the case of $\BC^3/\BZ_3$ orbifold without fixed points \cite{GRW},
 we get that $\bX/U(1) = \CP^2$. The curves of degree
one in $\CP^2$ are parametrized by a dual $\CP^2$. So in this case
we get $(N+1)(N+2)/2$ states. These can also be counted in the CFT.
Instead of having two fields $A_1,A_2$ to make the dibaryon, there are
 three such
fields, which transform in the $3$ of an $SU(3)$ global symmetry.
The totally symmetric combination of $N$ of these objects gives a
 Hilbert space of the
same dimension. Moreover there is an extra degeneracy by a factor
 of three from the
fact that the $S^5/\Z_3$ space is not simply connected, so there is a
possibility of
having discrete Wilson lines for the gauge field on the D-brane
\cite{GRW}. This factor of three is also reflected in the field
theory quiver
diagram, where there are three different types of dibaryon states.

Both of these examples show that the counting with $N$ times the hyperplane
bundle
of $\CP^m$ seems to give the right match for counting states.

With this in mind, let us return to the conifold case, and let us study a
D-brane that wraps once both of the two different $S^2$'s.
It was argued in \cite{BBNS} that such D-branes
should become giant gravitons. The counting of holomorphic
sections of the line bundle gives us $m = 3$. Indeed, if we consider
$x_1,x_2$ and $y_1,y_2$ a set of projective
coordinates for both of the $S^2$, the global sections of this line
bundle are given by curves
\begin{equation}\label{eq:cover1}
\sum a_{ij} x_i y_j = 0
\end{equation}
where we get four possible distinct coefficients $a_{ij}$.

The number of such states is therefore
${(N+1)(N+2)(N+3)}/{3!}$. This is many more states than just the
product of two dibaryons, one for $A$ and one for $B$. Indeed, these
factorized  states
correspond to the curves that factorize into linear terms in the
equation \ref{eq:cover1}. But when we have intersecting branes, we have
also deformations that are localized at the intersection.
These are the ones
responsible for giving us a larger moduli space.

In the case where the curve factorizes, the operator is given
by
\begin{equation}
\epsilon^1\epsilon_2 (A,\dots,A) \, \epsilon^2\epsilon_1(B,\dots ,B) \ .
\end{equation}
Notice now that the total baryon number\footnote{The baryon number
 counts the
number of $A$-fields minus the number of $B$ fields, with a choice of
normalization which we set equal to one for the smallest dibaryon operator.}
 of this operator is zero, and that
the $\epsilon^1$ appears as many times as the $\epsilon_1$.
We can use the identity
\begin{equation}
\epsilon^{a_1\dots a_N}\epsilon_{\tilde a_1 \dots \tilde a_N}
= \sum_{\sigma} (-1)^{|\sigma|} \delta^{a_1}_{\sigma(\tilde a_1)}\dots
\delta^{a_{N}} _{\sigma(\tilde a_N)}
\end{equation}
where $\sigma$ sums over all possible permutations of the $\tilde a_i$, and
$|\sigma|$ is $0$ or $1$ if the permutation is even or odd.

This procedure lets us eliminate one set of the $\epsilon$ symbols
if we want to. We see that if we do this, the $A$ have to be paired with the
$B$ in combinations of the form $AB$. There are $4$ such possible
combinations.
Consider therefore the operators of the form
\begin{equation}
\epsilon^1\epsilon_1((AB),(AB), \dots, (AB))
\end{equation}
where each of the entries is given by one of the four possible
combinations $A_iB_j$. This object can be expressed in terms of traces
of the $AB$, so it is built out of gravitons, and indeed
it should represent a maximal giant graviton.
We have a symmetric combination of four objects,
and the combinatorics of these objects also gives us  a total of
${N+3\choose 3}$ objects. In some sense, $A_i$ and $B_j$ are the analog of
the coordinates $x_i$ and $y_j$.

Similarly, if we take a state which wraps sphere one twice and sphere
two once, we get that  $m = 3\times 2 -1$. The natural objects to consider
are of the form $A_iB_j A_k$. However, we have to remember the $F$-terms, so in
this expression we get a field which is symmetric in $i,k$. The total
number of objects that we can count is $6$, and we can build operators
of the form
\begin{equation}
\epsilon^1 \epsilon_2( (ABA),(ABA), \dots, (ABA)) \ .
\end{equation}
Notice that this operator corresponds to
some form of a coherent state of excitations of the
dibaryon (\ref{eq:dibaryon}). {}From the supergravity,
we expect such a result
because this new baryon-like state
is wrapping the same topological cycle as the old dibaryon.

For more general wrapping cases of baryon number two or higher, it is more
difficult to match the states. Since we cannot get rid of all but one of the
pairs of $\epsilon$ symbols, it is not true that there is an easy
 description of the state in terms of $N$-symmetrized products
of similar objects (the $ABA$ above). It would be interesting to obtain
a more thorough description of these other states.

\section{BPS fluctuations of dibaryons}

In this section we show
that there exist BPS excitations of
the dibaryon operators, that is, operators which carry baryon number
and have higher conformal weight than the volume of the
associated 3-cycle.
Since in this particular section we will be interested in computing
explicitly the fluctuation spectra of D-branes, we need to resort
to writing models for
specific examples where the metric and the CFT
are both known. We are moreover interested in situations where we
have only ${\mathcal N}=1$ supersymmetry without singularities.
In these models the $R$ symmetry is strictly $U(1)$ and the
chiral fields are holomorphic.
Two of these models are particularly simple.
These are D-branes at the
 conifold \cite{KW} and D-branes at the orbifold $\BC^3/\BZ_3$
\cite{GRW}
without fixed points. We will deal in this section with
the particular case of the conifold.

%

Let us consider for simplicity the state with maximum $J_3$
of the first $SU(2)$:
\begin{equation}
\epsilon^1\epsilon_2(A_1, \dots , A_1)= {\rm det} A_1 \ .
\label{A1baryon}
\end{equation}
To construct excited dibaryons we can
replace one of the $A_1$ by any other chiral field which
transforms in the same representation of the gauge groups.
One possibility is to replace $A_1\rightarrow A_1 B_i A_j$,
where the two gauge indices of $B$ are contracted
separately with the
$A_1$ and
with the $A_j$, following the rules for matrix multiplication, namely
\begin{equation}
(A_1B_iA_j)^{b_1}_{a_2} = (A_1)^{b_1}_{a_1} (B_i)^{a_1}_{b_2} (A_j)^{b_2}_{a_2}
\ .
\end{equation}

This dibaryon-like state
is a chiral field up to F-terms because it is a gauge invariant holomorphic
polynomial in the chiral superfields.
There will be
a chiral primary state
with the same quantum numbers as the above operator that is
a linear combination of operators of this type.  Remember that operators
 that
differ by F-terms are equivalent as elements of the chiral ring, but
in the conformal field theory they are different and only one particular
linear combination is protected.

The operator resulting after the replacement
factorizes into the original dibaryon and
a single-trace operator:
\begin{equation} {\rm Tr} (B_i A_j)\ {\rm det} A_1 \ .
\label{factor}
\end{equation}
The factorization
suggests that this excitation of a dibaryon can be represented as
a graviton fluctuation in presence of the original dibaryon.

Not all excitations factorize in this way, however.
For example, consider replacing $A_1\rightarrow
A_2B_i A_2$.
One might ask whether
this new operator can be written as a product of the original dibaryon
$\det A_1$ and
a meson-like operator of the form $\tr(A_2 B_i)$.
 The answer is no.
To see why, it is
 easier to go to a generic point in the  moduli space of vacua of the theory
 where we can set
$A_1={\bf 1} a$
by gauge transformations, and thus we establish an isomorphism between the
indices of the two gauge groups. The operator above becomes
$a^{n-1} \tr(A_2B_iA_2)$, and if $N$ is large enough, there are no
relations amongst traces of a low number of fields.
In other words, we cannot write the operator as
$\epsilon^1\epsilon_2(A_1,\dots, A_1,A_2)\tr(B_iA_2)\sim
 a^{n-1}\tr(A_2)\tr(B_iA_2)$.
Since this new
operator cannot be factored it has to be interpreted as a single
particle state in AdS.\footnote{
Another way of seeing that this new operator is not a multiparticle
state begins with the observation that in general
\begin{equation}
\epsilon^1 \epsilon_2 (A_1, A_1, \dots , A_1, C)
= \frac{1}{N} \tr(A_1^{-1} C) \epsilon^1\epsilon_2(A_1, A_1, \dots , A_1)
\end{equation}
where $C$ is some operator with the same transformation properties
as the $A_i$.  From this formula, it is clear
that if $C=A_1B_iA_1$, $A_1B_iA_2$, or $A_2B_iA_1$, the new operator factors
inside the ring of chiral primaries.  However, for the case we picked
above, $C=A_2B_iA_2$, the $A_1^{-1}$ cannot be eliminated and the operator
does not factor.}
Since the operator also carries baryon number one,
the natural conclusion is that the one-particle state is
a BPS excitation of the wrapped D3-brane in the dual string theory.

Let us now study
BPS fluctuations of the wrapped D3-brane
in the supergravity approximation.
The DBI action is a good approximation
in the limit of weak string coupling and weak curvature of
the D-brane. These conditions are met in the
limit of large `t Hooft coupling.
In particular, we will compute the spectrum of quadratic fluctuations
of this DBI action.
We will return to  the field theory later to make the
correspondence of states and quantum numbers more precise.

First, we set up the
DBI computation in the right coordinate system.
The full ten dimensional metric is naturally $AdS_5 \times T^{1,1}$,
\begin{equation}
ds^2 = L^2 (-\cosh^2(\rho) d\tau^2 + d\rho^2 + \sinh^2(\rho) d\Omega_3^2) +
L^2 g \ .
\label{tenmet}
\end{equation}
For convenience, we have chosen the time $\tau$ direction to be a Killing
vector.  The radius of curvature is $L$.
The metric of $T^{1,1}$, the base of the conifold, is given by
\begin{equation}
g =b^2 \left[ A^2(d\psi +
\cos(\theta_1)d\phi_1 + \cos(\theta_2) d\phi_2)^2+ \sum_{i=1}^2
[d\theta_i^2+\sin^2(\theta_i)d\phi_i^2] \right]
\end{equation}
with $A^2 = 2/3 $, $b^2 = 1/6$.
We will keep $A,b$ as variables in our computation
for consistency checks.

The dibaryon is chosen to wrap
the cycle defined by $\theta_2, \phi_2$ constant.
This configuration is invariant under rotations of the
sphere wrapped by the D3-brane, but it is not invariant under the
$SU(2)$ associated to the $\theta_2,\phi_2$ coordinates.
The induced metric on the dibaryon is thus
\begin{equation}
(Lb)^{-2}g_{ind} =  -b^{-2} \cosh^2(\rho) d\tau^2 +
A^2(d\psi + \cos(\theta_1)d\phi_1)^2+
(d\theta_1^2+\sin^2(\theta_1)d\phi_1^2) \ .
\end{equation}
It is convenient to make a change of variables $\cos(\theta_1) = x$, so that
\begin{equation}
(Lb)^{-2}g_{ind} =  -b^{-2} \cosh^2(\rho) d\tau^2 +
A^2(d\psi + x d\phi_1)^2+\left[\frac{dx^2}{1-x^2}+(1-x^2)d\phi_1^2 \right] \ .
\end{equation}

In these variables, the determinant of
the spatial part of the metric
\begin{equation}
\frac{ -\det{g_{ind}} }{L^2 \cosh^2(\rho)} =(Lb)^6 A^2
\end{equation}
 is constant.
The variables' range is given by  $\psi\in [0,4\pi)$,
 $\phi_1\in[0,2\pi)$ and $x\in[-1,1]$.
The volume of the wrapped manifold is thus $L^3 16\pi^2 Ab^3= 8L^3\pi^2/9$.
In \cite{GK}, it was noted that this volume times the tension of
the D3-brane should be a good approximation of the mass of the
corresponding dibaryon.
As we argued in section 2.4, this volume is directly proportional
to the dimension $\Delta$ of the dibaryonic operator.
Indeed, as the conformal dimension
of each $A_i$ and $B_i$ is 3/4, one finds that
$\mu_3 8L^4\pi^2/9 = 3N/4$
is exactly equal to $\Delta$.
We return now to calculating the
excitation spectrum of the dibaryon.


We need to be careful with single valued functions on this space.
This squashed $S^3$ can be thought of,
essentially, as the group manifold $SU(2)$.  The coordinates
$(\psi,\theta_1,\phi_1)$ are the Euler angles.
One might have thought that the points $(\psi, \theta_1, \phi_1)$
and $(\psi, \theta_1, \phi_1+2\pi)$ were equivalent.
The insight from $SU(2)$ lets us correct this mistake.  Equivalent
points on $SU(2)$ have
\begin{equation}
\psi + \phi_1 \equiv \psi' + \phi_1' \mod 4 \pi \ ,
\end{equation}
\begin{equation}
\psi - \phi_1 \equiv \psi' - \phi_1' \mod 4 \pi \ ,
\label{singval}
\end{equation}
where $(\psi,\theta,\phi)$ and $(\psi',\theta',\phi')$
are two points on $SU(2)$.


We want to find the normal modes of oscillation of the wrapped D3-brane
around the solution corresponding to some fixed world-line in $AdS_5$ and some
fixed $\theta_2$ and $\phi_2$ on the transverse $S^2$.
The fluctuations along the transverse $S^2$ are the most interesting:
they change the $SU(2)\times SU(2)$ quantum numbers
and are most usefully compared with the chiral primary states
in the field theory.  The transverse fluctuations along the
$AdS_5$ are considered briefly afterward.
Supersymmetry relates the gauge field degrees of freedom
and fermions on the D-brane
to the scalar modes considered here.
There is no mixing between the between the different modes
at quadratic order in the fluctuations, and
we ignore the vector and spinor modes in what follows.

Because the fluctuations around $\phi_2$ and $\theta_2$ are by
definition small,
it is appropriate to treat the $S^2$ parametrized by these coordinates
as a flat $\R{2}$.
For example, one may take $\theta_2 \approx 0$.
Then the fluctuation coordinates on $S^2$ can
be taken to be
$y_1 = \theta_2 \sin(\phi_2)$ and $y_2 = \theta_2 \cos(\phi_2)$.
The connection term in the metric on $T^{1,1}$ becomes
$\cos(\theta_2) d\phi_2 = d\phi_2 +\frac{1}{2}(y_1 dy_2 - y_2 dy_1)$
and the K\"ahler form on the transverse
$S^2$ becomes $\sin(\theta_2) d\phi_2 \wedge d\theta_2 = dy_1 \wedge dy_2$.
To say the same thing in a different way, we are interested in the D-brane
action to quadratic order in fluctuations.
We only need to consider terms of up to order $y^2$, $y dy $, or
$dy^2$, and we neglect everything else which is higher order in the
fluctuations $y$.


Notice that the ten dimensional metric (\ref{tenmet}) has a term of the form
$(d\psi + A_\mu dx^\mu)^2$, where the $x^\mu$ are the $\phi_1$ and $\phi_2$
coordinates.  Indeed, in the previous section, we saw that we could add a more
general term of this form depending on the $AdS_5$ coordinates.
This perturbation corresponded to a gauge field carrying the $R$-charge.
Recall that changing the coordinate $\psi$ to
$\psi'=\psi+f(x)$ does not change the periodicity of the variable
$\psi'$; but it does change the form of the vector $A_\mu$ by a gauge
transformation, $A_\mu\to A_\mu -\partial_\mu f$. So in writing the metric,
the invariant quantity is the field strength of $A$.
We are in this way free to add a term of the form $-d\phi_2$ to $A$,
changing $d(\cos(\theta_2)d\phi_2) = d(\frac{1}{2}(y_1dy_2-y_2dy_1))$
and making the connection term simpler.


The Dirac-Born-Infeld acion is given by
\begin{equation}
S_{eff}  = -\mu_3  \int d^4x \,
\sqrt{-\det{g_{\mu\nu}
\partial_\alpha x^\mu\partial_\beta x^\nu }}
+ \mu_3 \int C_4
\end{equation}
with $\mu_3$ the tension of the brane.
Because $y_1, y_2$ appear only quadratically in the metric, we
can study the quadratic fluctuations in $y$ by doing
a first order variation in the metric
\begin{equation}
S_{eff} = S_0 -\frac{\mu_3}{2} \int d^4x \, \sqrt{-\det{g^{(0)}_{ind} }}
\ \tr \left[{g^{-1}_{ind }} \delta g_{ind} \right]
+ \mu_3 \int C_4
\end{equation}
where $\delta g_{ind}$ is the second order
contribution from the fluctuations in $y_1, y_2$
and we define $g^{-1}_{ind}$ to be the inverse of $g^{(0)}_{ind}$.
We can also do the same with the transverse fluctuations along the $AdS_5$,
but we will leave those for later.

With the coordinates chosen $\sqrt{-\det{g^{(0)}_{ind}}}$ is
independent of the fluctuations $y_1, y_2$ and also
independent of the D3-brane coordinates.
As a result, the wave equation on the D-brane worldvolume
is simplified.
We fix the diffeomorphism invariance of the
DBI action by locking the internal brane coordinates to the
background coordinates $\psi,\phi_1,x ,\tau$.
This locking corresponds to choosing a physical
gauge.
We take $\rho=0$.
If the brane is not moving then
\begin{equation}
L^{-2} g^{(0)}_{ind}= \begin{pmatrix}
A^2b^2&A^2b^2 x&0&0\\
A^2b^2 x& b^2(1-x^2)+A^2b^2 x^2&0&0\\
0&0&\frac{b^2}{1-x^2}&0\\
0&0&0&-1
\end{pmatrix}
\end{equation}
The inverse matrix is

\begin{equation}
L^2 g_{ind}^{-1}= \begin{pmatrix}
\frac{1-x^2+A^2 x^2}{A^2b^2(1-x^2)} &
-\frac{x}{b^2(1-x^2)}&0&0\\
 -\frac{x}{b^2(1-x^2)}&\frac{1}{b^2(1-x^2)}&0&0\\
0&0&\frac{1-x^2}{b^2}&0\\
0&0&0&-1
\end{pmatrix}
\end{equation}

We may choose a gauge where $C_4$ has a piece in
the $\psi$, $\theta_1$, $\phi_1$ and $\phi_2$ directions.
More specifically, we will choose a gauge which is well
defined at the north pole of the sphere parametrized
by $\theta_2$ and $\phi_2$:
\begin{equation}
C_4 = 4 Ab^5 L^4 \left(-1 + \cos(\theta_2) \right)
\sin(\theta_1)
d\psi \wedge d\theta_1 \wedge d\phi_1 \wedge d\phi_2  \ .
\end{equation}
We want to expand the probe brane action to quadratic order in fluctuations
in $y_1$ and $y_2$.  At this order, the four form $C_4$ is
approximately
\begin{equation}
C_4 \approx  -2b^2 \sqrt{ -\det{g^{(0)}_{ind}} } \, (y_1 dy_2 - y_2 dy_1)
d\psi \wedge dx \wedge d\phi_1
\ .
\label{Cfour}
\end{equation}

{}Putting the various pieces together, we obtain that
\begin{equation}
\tr{g_{ind}^{-1}\delta g}
\ = \sum_i g_{ind}^{\alpha\beta} L^2 b^2(\partial_\alpha y_i\partial_\beta y_i)
+2 g_{ind}^{\alpha\beta}{g_{\alpha i} \partial_\beta y^i} \ .
\end{equation}
The first term gives rise to
 the standard laplacian on the three-sphere, while the second term
gives rise to mixing terms between $y^1, y^2$ that are only first order in
derivatives.  An additional mixing term comes from (\ref{Cfour}).
These three contributions give
us the effective Lagrangian density to quadratic order in fluctuations.
\begin{equation}
{\mathcal L} = \tr g_{ind}^{-1} \delta g + 4b^2
(y_1 \partial_\tau y_2 - y_2 \partial_\tau y_1) \ .
\end{equation}

The non-trivial elements of $g_{\alpha i}$ are
\begin{equation}
g_{\psi i}=
-(1/2)L^2A^2b^2\epsilon_{ij}y^j  \; ; \; \; \; \;
g_{\phi_1 i}=-(1/2)L^2A^2b^2 x \epsilon_{ij}y^j \ .
\end{equation}
{}From now on, we will call $\phi_1$ simply $\phi$.
The spatial mixing term
$2 g_{ind}^{\alpha\beta} g_{\alpha i} \partial_\beta y^i$
is explicitly given by
\begin{equation*}
-\frac{L^2}{2} \left[ g^{\psi\psi} A^2b^2 y^2 \partial_\psi y^1
+g^{\psi\phi} A^2b^2 y^2 \partial_\phi y^1
+g^{\phi\psi} A^2b^2 x y^2 \partial_\psi y^1 \right.
\end{equation*}
\begin{equation}
\left. +g^{\phi\phi} A^2 b^2 x y^2 \partial_\phi y^1
-(1\leftrightarrow 2)\right] \ .
\end{equation}

The equations of motion for the fluctuations are given by
\begin{equation}
L^2b^2 \partial_\alpha (g_{ind}^{\alpha\beta}\partial_\beta y^i)
+\partial_\beta(g_{ind}^{\alpha\beta} g_{\alpha i})
-(\partial_{y_i} g_{\alpha i}) g_{ind}^{\alpha\beta} \partial_\beta y^i
-4b^2 \epsilon^{ij} \partial_\tau y^j
= 0 \ .
\end{equation}
Notice that the elements of $g_{ind}^{\alpha\beta}$ are independent of the
variables with respect to which we are taking derivatives.
As a result, the second and third terms in the equation above
are actually equal and we have
\begin{equation}
L^2b^2 \partial_\alpha (g_{ind}^{\alpha\beta}\partial_\beta y^i)
+2\partial_\beta(g_{ind}^{\alpha\beta} g_{\alpha i})
- 4 b^2 \epsilon^{ij} \partial_\tau y^j
= 0 \ .
\end{equation}
Taking the combinations $y^{\pm}= y^1\pm i y^2$ these equations become
\begin{equation*}
b^2(\nabla^2 y^{\pm}-L^{-2} \partial_\tau^2 y^{\pm})
\mp i4b^2 L^{-2} \partial_\tau y^{\pm}
\end{equation*}
\begin{equation}
\pm i b^2A^2 (g^{\psi\psi} \partial_\psi y^{\pm} +g^{\psi\phi}
x \partial_\psi y^{\pm} +g^{\psi\phi}   \partial_\phi y^{\pm}
+g^{\phi\phi} x\partial_\phi y^{\pm}) =0 \ .
\end{equation}
Now, we see that $g^{\psi\phi}= - x g^{\phi\phi}$ so the terms with
derivatives with respect to $\phi$ cancel. This is just as expected, since
the result should be invariant under the $SU(2)$ of isometries
of the squashed $S^3$. Thus far we are getting a consistent picture.

The terms with $\partial_\psi y^{\pm}$ are given by the combination
\begin{equation}
g^{\psi\psi} +g^{\psi\phi} x= \frac{1}{L^2A^2b^2}
\end{equation}
which is a constant coefficient.

Now we can use the separation of variables
\begin{equation}
y^{\pm} = \exp(-i\omega \tau) \exp(im \psi)\exp(i n \phi) Y^{k\pm}_{mn}(x)
\end{equation}
to obtain a differential  equation for $Y^{k\pm}_{mn}(x)$.
Note that for the $y^\pm$ to be single valued,
the condition (\ref{singval}) implies that $m$ and $n$ are
either both integer or both half-integer.
The remaining differential equation for the
$Y^{k\pm}_{mn}$ is given by
\begin{eqnarray}
\lefteqn{\partial_x (1-x^2) \partial_x Y^{k\pm}_{mn} =}
\label{hgeom}
\\
&& b^2\left(\frac{1-x^2+A^2 x^2}{A^2b^2(1-x^2)}
m^2 -\frac{2x}{b^2(1-x^2)}mn + \frac{n^2}{b^2(1-x^2)}-\omega(\omega \pm 4)
\pm b^{-2}m
 \right) Y^{k\pm}_{mn} \nonumber
\end{eqnarray}
Let us begin by analyzing
the behavior of the solution in the limit $x\to \infty$.
In this limit, the differential equation becomes
\begin{equation}
b^2 \left( \frac{1-A^2}{(A^2b^2)}m^2-\omega(\omega \pm 4)\pm b^{-2}m \right) Y^{k\pm}_{mn} +
\partial_x(x^2 \partial_x Y^{k\pm}_{mn} ) = 0 \ .
\end{equation}
The solution to this equation is clearly a power of $x$,
$Y^{k\pm}_{mn} \sim a_kx^k$.  Thus
\begin{equation}
\omega(\omega \pm 4)=
b^{-2} \left(k(k+1)- m(m\mp 1) \right)+\frac{m^2}{A^2b^2} \ .
\end{equation}
The energy should be real for all allowed values of $m$ and $k$.
Notice that the energy has the
right dependence in terms of the $SU(2)$ quantum numbers to be associated with
the velocity on a group manifold.

Only certain values of $k$ are allowed because
the fluctuations must be well-behaved at $x=\pm 1$.
The differential equation (\ref{hgeom}) can be solved in terms
of a hypergeometric
function
\begin{equation}
Y_{mn}^{k\pm} = (1-x)^\delta (1+x)^\epsilon
F(\alpha, \beta, \gamma; z=(x+1)/2) \ ,
\label{hgeomsolution}
\end{equation}
where $\alpha$, $\beta$, $\gamma$, $\delta$, and $\epsilon$
depend on the quantum numbers $k$, $m$, and $n$.
The north and south pole of the $S^2$, $x=\pm 1$,
 correspond to the singular points $z=0$ and $z=1$
of the hypergeometric function.
For the fluctuations to vanish at $x=\pm 1$,
$k-m$ must be a non-negative integer.\footnote{
The word ``must'' is a little too strong here.
In fact, a certain symmetry in the
functional form of (\ref{hgeomsolution}) means that
distinct choices of $k,m,n$ may correspond
to the same fluctuation.  To be more precise,
the choice $k - \max(m,n) \in \{0,1,2,\ldots\}$
and $m+n>0$ corresponds to non-singular fluctuations with
no redundancy.}

For the $y^\pm$
which contribute to the $U(1)$ charge in the same direction as the
unexcited D3-brane, the choice $k=m$ corresponds to a BPS state;
we have
\begin{equation}
\omega = m/Ab = 3m \ .
\end{equation}
Indeed, the BPS states should have the lowest possible dimension for a
given R-charge.  The states $k=m$ meet this condition.
{}From the periodicity of $\psi$, $m$ can be a half integer, and when
$m$ is a half integer so is $k$.

This energy spectrum
means that the contribution of these BPS states to the energy is quantized
in units of $3/2 L^{-1}$.
 $3/2$ is also exactly the change in the conformal dimension of the
chiral operators once $A_2 B_i A_2$ is substituted for an $A_1$ in
the antisymmetric product (\ref{A1baryon}).
These modes match the result from conformal
field theory.\footnote{$(AB)$ contributes $3/2 \times 2/3$
to the R-charge of the state.}

 Also, the transformation under
the $SU(2)$ that rotates $B_1,B_2$ is in agreement. The unexcited
dibaryon was a singlet while the excited one acquires spin $k=m$.
Indeed, for each value of $m$ there is a unique irreducible
representation of $SU(2)_B$ associated to it.
This means that the quantum of $U(1)$ charge proportional to $m$
should be associated to a spin $m$ state for $SU(2)_B$.
In the conformal field theory, these states result from
choosing to replace one of the $A_1$ in the $\epsilon^1\epsilon_2()$
function by, for example,
$A_2B_iA_2B_k \dots A_2$, where we have $2m$ $B$'s inside the matrix.
Using the F-term equations of motion, it can  be seen that these states are
totally symmetric with respect to exchange of the $B$ variables.
Since the $B_{1,2}$
carry spin $1/2$ under the $SU(2)_B$, these states with $2m$ $B$'s form a
totally symmetric
representation of $SU(2)_B$ with spin $m$. Therefore, we
can match the $SU(2)_B$ quantum numbers of the states to the supergravity.

We have not, however, determined the transformation property of the excited
dibaryons under the $SU(2)$ that rotates $A_1,A_2$. This is more difficult
since we need to consider the coupling of the fluctuation fields
with the zero-mode dynamics.
Since we have not determined the full $SU(2)\times SU(2)$ quantum
numbers, we cannot make a precise determination of the dual
gauge theory operators. In particular, it would be interesting
to see if the factorized operators of the type (\ref{factor})
are necessary to match the spectrum of the fluctuating probe D3-brane.
We leave this interesting question for the future.

For the transverse motion in $AdS_5$ there is no mixing of the directions,
 so we
get four scalars and their energies are given by
\begin{equation}
\omega^2 = b^{-2} (k(k+1)- m^2)+ \frac{m^2}{A^2b^2} + M^2
\end{equation}
with $M$ the mass of these states. This mass term comes because
the space-time is curved and to second order
$g_{00} = -L^2 (1 + \rho^2 + \ldots)$; the AdS excitation feels
a gravitational potential.
For $k=m$ the above expression should be a perfect square for all $m$
(it is a superpartner of the other BPS states), so
we should have $M^2 A^2 b^2 = A^4/4$, or equivalently
$M^2 = A^2/4b^2=1$.  This relation
can be checked explicitly from the metric.
It follows that
$ \omega = 3 m +3 A^2/2 = 3m +1$ differs by $1$ from the previous
energy.  Indeed, we expect excitations in the AdS directions
to correspond to introducing covariant derivatives
for the fields $A_i$ or $A_2B_iA_2$ on the field theory side.
Covariant derivatives have conformal weight one.
This is an alternative check that the normalization of the $C$ field
we chose is correct, since it predicts that the
splitting between the
energies of the modes is compatible with the spacetime
symmetries.

Notice that the dibaryon state on the gravity side has a Fock space
worth of excitations.
In the field theory, this Fock space can be reproduced as
well. To insert one quantum, we took one of the $A_1$ and replaced it with
$A_2BA_2B\dots$. To put many quanta, we replace many of the $A_1$ by the
$A_2BA_2BA_2\dots$ combinations. The fact that we have a Fock space
of identical particles, as opposed to distinguishable particles,
comes from the permutation symmetry of the $\epsilon$ symbols. For fermionic
excitations we need to remember
the $(-)$ signs when we exchange the fermionic
insertions in the operator.

We conclude that we can match, at least schematically, the dibaryon state
 with any number of BPS open strings that excite it.
We have not carried out the complete matching since we have not
determined the $SU(2)_A$ quantum numbers of the excited D3-branes,
which remains an interesting problem for future work.
We believe that the construction of non-BPS states can be carried out
using similar methods to the ones used in \cite{BMN} for closed
strings, but this construction is beyond the scope of the present paper.

\section{Discussion}

In this paper we studied in some detail the correspondence between
D3-branes wrapping 3-spheres inside $T^{1,1}$
and baryon-type operators
of the dual $SU(N)\times SU(N)$ gauge theory. By calculating
the $U(1)_R$ charge of the D3-branes and their collective coordinate
energies we have new evidence that the operator identification
proposed in \cite{GK} is correct. Furthermore, we showed that
there exist BPS excitations of wrapped D3-branes and suggested the chiral
operators dual to them.
Our results provide new evidence that the duality between gauge invariant
operators of a superconformal gauge theory and states of string
theory on $AdS_5\times \bX_5$ extends to operators whose dimensions grow
as $N$, the number of colors. It is clear that this same procedure to generate
operators should work in other examples related to different quiver
theories and that one can generically find BPS excitations of baryon-like
operators.

There is another such class of operators, similar to the ones we have
considered. It is related to the ``giant graviton'' effect \cite{MST}.
It has been observed that for modes whose angular momentum on $\bX_5$
is of order $N$ the single-trace description of the dual gauge theory
operators breaks down \cite{BBNS,CJR}. Instead, the correct description is in
terms of subdeterminants of elementary fields, which for maximum angular
momentum become determinants similar to the dibaryon operators we have
considered. On the string theory side, the modes whose angular
momentum is of order $N$ blow up into D3-branes on
$\bX_5$.\footnote{These
objects are not topologically stable. The blow-up happens for dynamical
reasons.}
This is another manifestation of the fact that D-branes, and therefore
string theory, are crucial for
describing gauge invariant operators whose dimensions grow as $N$.
Thus, string theory is necessary to complete the state/operator
map even at large `t Hooft coupling. Along these lines, it is
interesting to consider the BPS excitations of the giant gravitons
and to construct the dual operators. Some results on this
are presented in \cite{DJM}. We believe that the giant graviton case is
similar to our study of topologically stable wrapped D3-branes.

Very recently, a much more dramatic demonstration of the stringy
nature of the AdS/CFT duality at large `t Hooft coupling was
presented in \cite{BMN}. The insight of this paper is to focus on states
whose angular momentum $J$ on $S^5$ scales as $\sqrt{N}$. It was shown
that there exists a class of operators whose $\Delta- J$ stays
finite in this limit. These states are in one-to-one correspondence with
{\it all} the closed string states on a RR-charged pp-wave background,
including all the massive string states.
We believe that the work we have presented on the BPS excitations of
D3-branes constitutes a first step towards studying the open string
states in a similar setting. An unexcited D3-brane is described by
the basic dibaryon operators (\ref{BaryonOne}) and (\ref{BaryonTwo}).
Thus, these operators describe the open string vacuum. The more complicated
operators discussed in section 4 correspond to the BPS states of the
open string.
The non-BPS open string states can be constructed along
the lines of \cite{BMN}, but this construction is
beyond the scope of the present paper. The basic issue in question is
the proper understanding of the non-planar diagrams. When
the operator dimension scales as $N$, it is
too high for the planar approximation to be
valid.

\section*{Acknowledgements}
We are grateful to V.~Balasubramanian,
C.~Beasley, A.~Bergman,
A.~Hashimoto,
B.~Kol, J.~Maldacena,
P.~Ouyang,
M.~Strassler
 and E.~Witten
for useful discussions. This research was supported in part by the NSF Grants
PHY-9802484 and PHY-0070928, and DOE grant DE-FG02-90ER40542.

\end{document}